\documentclass{emulateapj}
\usepackage{amssymb, amsmath}
\usepackage{natbib}
\usepackage{graphicx}

\begin{document}
\title{Statistical Tools for Classifying Galaxy Group Dynamics}
\author{Annie Hou}
\affil{Department of Physics $\&$ Astronomy, McMaster University, Hamilton ON L8S 4M1, Canada}
\email{houa2@physics.mcmaster.ca}
\author{Laura C. Parker}
\affil{Department of Physics $\&$ Astronomy, McMaster University, Hamilton ON L8S 4M1, Canada}
\email{lparker@physics.mcmaster.ca}
\author{William E. Harris}
\affil{Department of Physics $\&$ Astronomy, McMaster University, Hamilton ON L8S 4M1, Canada}
\email{harris@physics.mcmaster.ca}
\and
\author{David J. Wilman}
\affil{Max-Planck-Institut f$\ddot{u}$r Extraterrestrische Physik, Giessenbachstra$\ss$e,  D-85748 Garching, Germany}
\email{dwilman@mpe.mpg.de}
\begin{abstract}
The dynamical state of galaxy groups at intermediate redshifts can provide information about the growth of structure in the universe.  We examine three goodness-of-fit tests, the Anderson--Darling (A--D), Kolmogorov and $\chi^{2}$ tests, in order to determine which statistical tool is best able to distinguish between groups that are relaxed and those that are dynamically complex.  We perform Monte Carlo simulations of these three tests and show that the $\chi^{2}$ test is profoundly unreliable for groups with fewer than 30 members.  Power studies of the Kolmogorov and A--D tests are conducted to test their robustness for various sample sizes.  We then apply these tests to a sample of the second Canadian Network for Observational Cosmology Redshift Survey (CNOC2) galaxy groups and find that the A--D test is far more reliable and powerful at detecting real departures from an underlying Gaussian distribution than the more commonly used $\chi^{2}$ and Kolmogorov tests.  We use this statistic to classify a sample of the CNOC2 groups and find that 34 of 106 groups are inconsistent with an underlying Gaussian velocity distribution, and thus do not appear relaxed.  In addition, we compute velocity dispersion profiles (VDPs) for all groups with more than 20 members and compare the overall features of the Gaussian and non-Gaussian groups, finding that the VDPs of the non-Gaussian groups are distinct from those classified as Gaussian.
\end{abstract}

\section{Introduction}
\indent{}The group environment represents an intermediate size and density scale between individual galaxies and rich galaxy clusters.  With roughly half of the present-day galaxy population in groups \citep{1983ApJS...52...61G, 2005MNRAS.362.1233E}, this environment plays an important role in the evolution of galaxies.  The relatively low velocity dispersions of groups provides an ideal environment for galaxy interactions \citep{1985MNRAS.215..517B,1998ApJ...496...39Z,2006MNRAS.370.1223B} and studying the dynamics of galaxy groups is one way to probe how dominant these interactions are in the group environment.\\
\indent{}The dynamics of rich clusters are often studied using extended X-ray emission, which provides information about the potential well of the cluster.  Since not all groups are sufficiently massive and evolved that their X-ray emission can be detected \citep{1998ApJ...496...39Z,2000ARA&A..38..289M}, we must use another tracer of galactic dynamics.  Another method of characterizing the dynamical state of gravitating systems such as galaxy groups, clusters, cores in molecular clouds, or star clusters involves studying their \emph{velocity distributions}.  The standard assumption is that the underlying distribution is Gaussian in nature, but this is strictly true only for systems in dynamical equilibrium.  Groups with non-Gaussian velocity distributions could mark systems in the process of a merger \citep{1996ApJ...472...46M} or those that are in the early stages of evolution.  However, the interpretation of galaxy group dynamics can be further complicated by projection effects \citep{1989ApJ...344...57R},  the possible inclusion of interloping galaxies \citep{2008ApJ...672..834R} and the direction of group elongation relative to the line of sight \citep{2009ApJ...696.1441T}.  To properly study group (or cluster) dynamics we need a reliable method of distinguishing between relaxed systems with Gaussian velocity distributions and more complex systems with non-Gaussian dynamics.\\
\indent{}Previous analyses of cluster velocity distributions have resulted in conflicting views on the dynamics of these systems.  \citet{1977ApJ...214..347Y} used the \emph{a}-test, \emph{u}-test and Shapiro--Wilks \emph{W}-test for non-normality to show that the observed radial velocity distributions of clusters of galaxies, with as few as 10 and as many as 122 members, are always consistent with an underlying Gaussian distribution.  More recently, evidence of substructure has been found in clusters \citep{1988AJ.....95..985D, 1994AJ....107.1637B,1998Sci...280..400B}, indicating that cluster dynamics may be more complicated than initially assumed.  \\
\indent{}\citet{1990AJ....100...32B} emphasize the difficulty in determining that a given velocity distribution differs significantly from Gaussian, stating that the goodness-of-fit tests used by \citet{1977ApJ...214..347Y} detect different departures (e.g., skewed or shifted distributions) from a true Gaussian distribution. Thus, a system may be classified as either Gaussian or non-Gaussian depending on the statistic used.  These difficulties are more severe when studying smaller systems, as in the case with galaxy groups.  Since group membership can range from 3 to 50 or more galaxies, we need a statistical test that is reliable even for extremely small sample sizes.  We also require a test that is robust, or unaffected by small departures from normality, to ensure that the rejections are a result of real deviations from a Gaussian distribution and not sensitivities inherent to the test.  The goodness-of-fit tests used to analyze rich clusters are generally not applicable to groups, where the challenges of small number statistics become relevant.  \\
\indent{}In this paper we test three goodness-of-fit techniques in order to determine which one in particular is best able to distinguish between Gaussian and non-Gaussian velocity distributions, especially for small samples.  In Section 2, we discuss the statistical tests we use to determine departures from Gaussianity and present the results of Monte Carlo simulations of the $\chi^{2}$, Kolmogorov and Anderson--Darling (A--D) tests, as well as the results of our power studies of the Kolmogorov and A--D tests.  In Section 3, we apply the goodness-of-fit tests to the Canadian Network for Observational Cosmology (CNOC2) galaxy group data and compare the results.  In Section 4 we apply the A--D test to the CNOC2 groups and classify the dynamical states of the groups.  In Section 5, we compare velocity dispersion profiles (VDPs) of the classified Gaussian and non-Gaussian galaxy groups and discuss the implications of our results and in $\S$6, we summarize our results\\
\indent{}Throughout this paper, we assume a $\Lambda$CDM cosmology with $\Omega_{M} = 0.3$, $\Omega_{\Lambda} = 0.7$ and $H_{0} = 75$ km s$^{-1}$ Mpc$^{-1}$.

\section{Statistical Tools}
\subsection{Goodness-of-Fits Tests}
\indent{}The Pearson's $\chi^{2}$ test (Equation \ref{chi}) is arguably the most commonly used goodness-of-fit test.  However, this statistic was developed as a large sample theory and its reliability begins to break down as one approaches small sample sizes (\emph{n}).  For small \emph{n}, \citet{D'agostino} (hereafter DA86) suggest the use of goodness-of-fits tests based on empirical distribution functions (EDFs), such as the A--D and the Kolmogorov tests.\\
\begin{center}
a.) Pearson's $\chi^{2}$ Test
\end{center}
\indent{}The Pearson's $\chi^{2}$ test is defined as:
\begin{center}
\begin{equation}
\chi^{2} = \sum_{i=1}^{n}\frac{(\mbox{observed$_{i}$ - expected$_{i}$})^{2}}{\mbox{expected$_{i}$}}
\label{chi}
\end{equation}
\end{center}
\paragraph*{}The critical points of the $\chi^{2}$ test were developed for large samples and it is difficult to determine exactly where, at small \emph{n}, this theory begins to break down (DA86).  Although there is no standard method to apply the $\chi^{2}$ test, it is often suggested that the bin widths be constructed such that they have the same number of data points in each bin.  For small sample sizes \citet{Vessereau} claims that it is not the number of data points per bin that is important but rather the \emph{total} number of data points.  \citet{Vessereau} also finds that as long as $n \geq 10$ and one uses the $1\%$ or $5\%$ critical values, the $\chi^{2}$ test should not produce a significant amount of error (with respect to false positives or negatives).  \citet{Koehler} claim that for \emph{k-1} degrees of freedom, the Pearson's $\chi^{2}$ test is `reasonably adequate' when $k\geq3$ and $n\geq10$ and \cite{Roscoe} also find that when the degrees of freedom are $> 1$, the $\chi^{2}$ test does not produce significant false positives, remaining robust against these type of errors.\\
\indent{}The $\chi^{2}$ test is often performed on binned data, testing the variance between histograms and a continuous Gaussian distribution.  Unfortunately, binning can add additional errors to the test, a problem which becomes worse at small \emph{n}.  One source of error is the choice of bin width, which can alter the results of the $\chi^{2}$ test, even causing the same data to be classified as both Gaussian and non-Gaussian, depending on the bin width.  Several authors \citep[e.g.][]{Heald,Scott} have attempted to reduce this effect by minimizing the sampling fluctuations.  With testing, Heald's optimal bin width was selected, where the bin width is defined as:

\begin{center}
\begin{equation}
\delta x  = \sigma\left(\frac{20}{n}\right)^{1/5}
\label{healdwidth}
\end{equation}
\end{center}
where $\sigma$ is the standard deviation.\\
\indent{}The degrees of freedom in the system are given by $DOF = N - 1 - k$, where $N$ is the number of bins and $k$ is the number of free parameters in the assumed distribution.  Thus, for a Gaussian distribution there are two free parameters ($k = 2$), $\mu$ and $\sigma$, and the minimum number of bins is $N=4$.\\

\begin{center}
b.) The Kolmogorov Test
\end{center}

\indent{}The popular Kolmogorov test (DA86, and references therein) is a goodness-of-fit test based on supremum statistics, which measures the vertical difference between the EDF, $F_{n}(x)$, of the \emph{ordered} data $x_{i}$ and the cumulative distribution function (CDF), $F(x)$, of a given model $\footnote{It should be noted that the Kolmogorov--Smirnov (K--S) test is technically the difference between two EDFs, while the Kolmogorov test measures the difference between a model CDF and EDF.  Despite this distinction, the Kolmogorov test is often referred to as the ``K--S test'', here we will refer to it by it's proper name.}$. The EDF statistic computed for the Kolmogorov test is the \emph{D} value, which is derived from the $\emph{D}^{+}$ and $\emph{D}^{-}$ values, and is defined as:
\begin{center}
\begin{equation}
D = max(D^{+},D^{-})
\label{Dvalue}
\end{equation}
\begin{equation}
D^{+} = \mbox{supremum}|\frac{i}{n} - F(x)|,
\end{equation}
\begin{equation}
D^{-} = \mbox{supremum}|F(x) - \frac{(i-1)}{n}|,
\end{equation}
\end{center}
where $F_{n}(x) = \frac{i}{n}$ for $D^{+}$, $F_{n}(x) = \frac{(i-1)}{n}$ for $D^{-}$, and $1 \leq i \leq n$.
\indent{}\citet{Stephens} has simplified the Kolmogorov test with the modification of the \emph{D} values, called the $D^{*}$ value (Equation \ref{Dstar}), which allows for comparison with one critical value table, rather than computing critical values for specific sample sizes and significance levels \citep{Massey}.\\
\begin{center}
\begin{equation}
D^{*} = D\left(\sqrt{n} + 0.12 + \frac{0.11}{\sqrt{n}}\right)
\label{Dstar}
\end{equation}
\end{center}

\indent{}The use of the Kolmogorov test in place of the $\chi^{2}$ test for small samples is suggested by \citet{Lilliefors}.  In a comparison of the Kolmogorov and $\chi^{2}$ tests, \citet{Massey} concludes that the Kolmogorov test is generally more reliable than the $\chi^{2}$ test, especially for small \emph{n}, where the effects of binning, required by the $\chi^{2}$ test but not the Kolmogorov test, can result in a large loss of information.

\begin{center}
c.) The Anderson--Darling Test
\end{center}
\indent{}Like the Kolmogorov test, the A--D test is also based on EDF statistics and does not require binning or graphical analysis.  Despite these advantages, the A--D test is not commonly used in astronomy.  The A--D statistic involves the calculation of the $A^{2}$ and $A^{2*}$ values, starting from the \emph{ordered} data $\{x_{i}\}$: 

\begin{center}
\begin{equation}
A^{2}= -n-\frac{1}{n}\sum_{i=1}^n(2i-1)(\ln\Phi(x_{i}) + \ln(1 - \Phi(x_{n+1-i}))),
\label{ADtest}
\end{equation}
\end{center}
\begin{center}
\begin{equation}
A^{2*} = A^{2}\left(1 + \frac{0.75}{n} + \frac{2.25}{n^{2}}\right)
\label{A2star}
\end{equation}
\end{center}
where $x_{i} \leq x < x_{i+1}$ and $\Phi(x_{i})$ is the CDF of the hypothetical underlying distribution.  The use of either $A^{2}$  or $A^{2*}$ for the A--D statistics depends on how well the distribution parameters are known \emph{a priori}.  The $A^{2*}$ values are modifications for cases where the distribution parameters are not known \emph{a priori} and must be estimated from the $x_{i}$ values.  If the input parameters are known beforehand, then the $A^{2}$ values should be used for comparison with critical value tables.  From Equation (\ref{A2star}), it is clear that $A^{2*}$ approaches $A^{2}$ for large \emph{n}.\\
\indent{}In this analysis, we take $\Phi(x_{i})$ to be the CDF of a Gaussian distribution, given as:

\begin{center}
 \begin{equation}
 \Phi(x_{i}) = \frac{1}{2}\left(1 + \mbox{erf}\left(\frac{x_{i} - \mu}{\sqrt{2}\sigma}\right)\right)
 \label{CDF}
 \end{equation}
\end{center}
where $x_{i}$ is the radial velocity of the galaxy group members, arranged from lowest to highest, $\mu$ is the computed mean velocity of the group and $\sigma$ is the computed velocity dispersion.  The $\Phi(x_{i})$ values are then used in the A--D computing formulas (Equations (\ref{ADtest}) and (\ref{A2star})) to obtain the $A^{2*}$ values.  These values can either be compared to known critical or limiting value tables, or used to compute the significance level $\alpha$, which gives the probability of the null hypothesis (i.e., the underlying distribution is Gaussian) being true.  A more detailed discussion of critical values and significance levels is given in Section 4.  DA86 recommend the A--D test as the `omnibus' test for EDF statistics when the underlying distribution is believed to be Gaussian.  Furthermore, they claim that the $A^{2}$ and $A^{2*}$ values can be reliably computed down to $n = 5$.\\
\indent{}A method of quantifying the robustness of statistical tests involves conducting power studies, which investigate the percentage of false positives a given test may produce when the underlying distribution is distorted (i.e., skewed, shifted, wings, etc.).  For the Gaussian distribution, this involves applying the tests to a variety of non-Gaussian samples and determining not only how often a specified test will reject the distribution as Gaussian, but also the specific types of departures from non-normality that affect the rejection rate (DA86, and references therein).  In a comparison of the Kolmogorov, Cram$\acute{e}$r-von Mises, Kuiper, Watson and A--D tests, \citet{Stephens} conducted power studies using a variety of non-Gaussian distributions and found the A--D test to be most powerful of the EDF statistics for detecting departures from Gaussianity, while the Kolmogorov test proved to be least powerful.  As for the $\chi^{2}$ test, DA86 claim that it is in general not a powerful test for Gaussian distributions and do not recommend its use.  We discuss the results of our power studies in Section 2.3.\\

% determining the power, which is defined as the probability of rejecting the null hypothesis (i.e. a sample is not derived from a given distribution) when the sample should in fact be rejected \citep{Zar}.  Power studies investigate the percentage of false positives a given test may produce given differing sample sizes.  For the Gaussian distribution, this involves applying the tests to a variety of non-Gaussian samples and determining not only how often a specified test will reject the distribution as Gaussian, but also the specific types of departures from non-normality that affect the rejection rate (DA86 and refs. therein).  Results of a power study conducted on EDF statistics by \citet{Stephens} suggest that the A--D test is the most powerful EDF test for Gaussianity, while the K test proved to be the least powerful.  As for the $\chi^{2}$ test, DA86 claim that it is in general not a powerful test for Gaussian distributions and do not recommend its use.
\subsection{Monte Carlo Simulations}
\begin{figure}[hbt!]
\includegraphics[width=8cm, height=8cm]{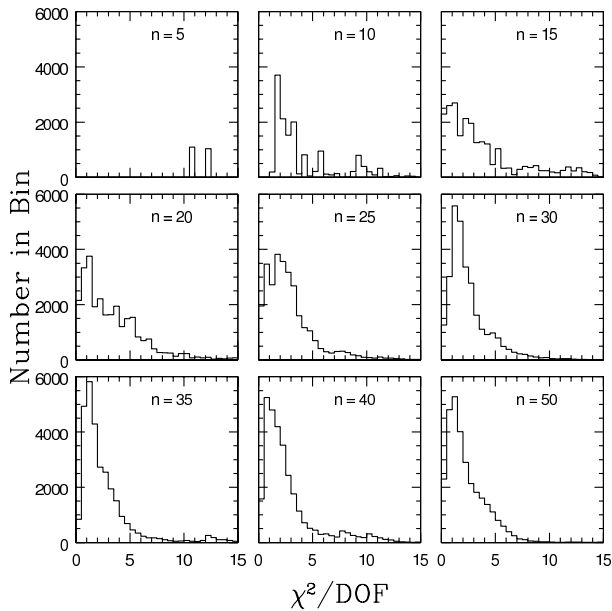}
\caption{Monte Carlo simulations of the $\chi^{2}$ test for different sample sizes, using a Gaussian random number generator with input  values of $\mu=0.0$ and $\sigma_{\rm{int}} = 1.0$, and 30 000 iterations.  The histograms are generated with a bin width of 0.5.  For $n <$ 30, the $\chi^{2}$ test completely fails to recover the input distribution.  This is most obvious for the $n = 5$ case, where the histogram shows two peaks at $\chi^{2}/DoF = 10,12$, instead of the expected peak value of $\chi^{2}/DoF = 1.0$.  Only for $n > 30$ do the simulations recover the expected peak value.}
\label{mcchi}
\end{figure}

\begin{figure}[hbt!]
\includegraphics[width=8cm,height=8cm]{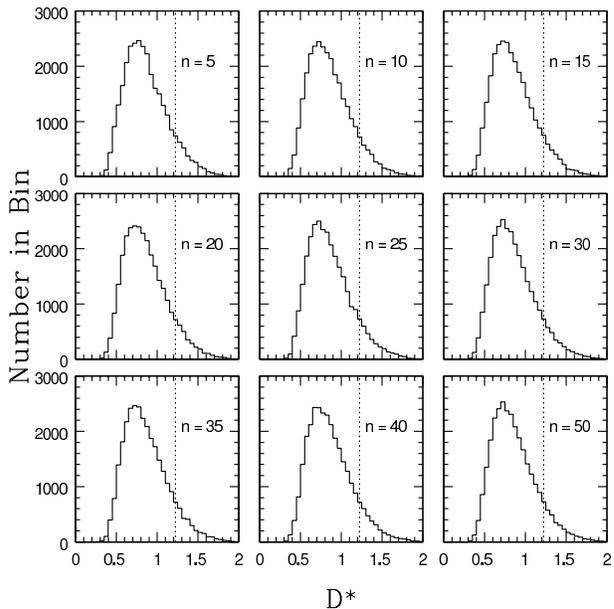}
\caption{Monte Carlo simulations of Kolmogorov test for different sample sizes, using a Gaussian random number generator with $\mu=0.0$ and $\sigma_{\rm{int}} = 1.0$, and 30,000 iterations.  The histograms of the $D^{*}$ values are plotted using a bin width of 0.05.  The results of the simulations show that even at $n = 5$, the Kolmogorov test is able to recover the input distribution.  The histograms for $n = 5$ to 50, consistently reproduce the expected peak values, indicating that the test is reliable over a wide sample range.  The vertical dotted line represents the $D^{*}$ value above which 10$\%$ of the values lie, our computed values are in complete agreement with the known $\alpha$ = 0.10 critical value of 1.224 (DA86).}
\label{mcKS}
\end{figure}

\begin{figure}[hbt!]
\includegraphics[width=8cm,height=8cm]{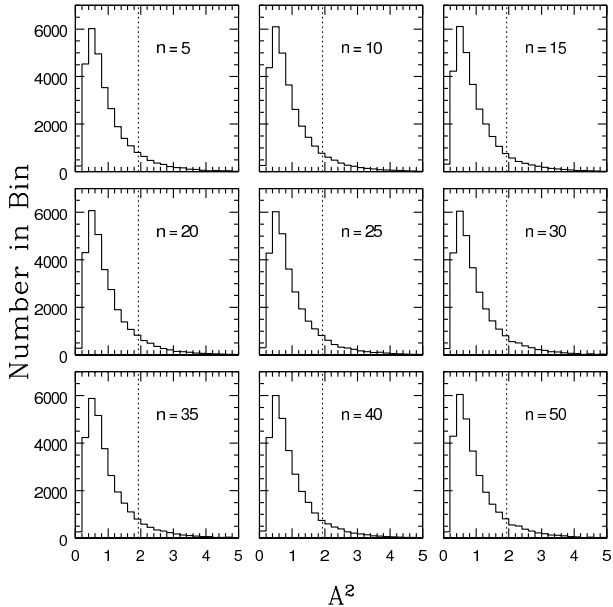}
\caption{Monte Carlo simulations of A--D Test for different sample sizes, using a Gaussian random number generator with input  values of $\mu=0.0$ and $\sigma_{\rm{int}} = 1.0$, and 30,000 iterations.  The histograms of the $A^{2}$ values are plotted using a bin width of 0.2.  The results of the simulations show that even at $n = 5$, the A--D test is able to recover the input distribution.  The histograms for $n = 5$ to 50, consistently reproduce the expected peak values, indicating that the test is reliable over a wide sample range.  The vertical dotted line represents the $A^{2}$ value above which 10$\%$ of the values lie, our computed values are in complete agreement with the known $\alpha$ = 0.10 critical value of 1.933 (DA86).}
\label{mcAD}
\end{figure}

\indent{}We test the claim of \citet{Stephens} and DA86 that the A--D is the most reliable test for Gaussianity for small sample sizes, by performing Monte Carlo simulations of the $\chi^{2}$, Kolmogorov, and A--D tests.  We perform 30,000 iterations for a variety of sample sizes, with $5\leq n \leq 50$ and drawing from a random Gaussian distribution with input values of $\mu = 0$ and $\sigma = 1.0$, to determine the reliability of the tests and how accurately each test can reproduce published critical values.\\
\indent{}The results of the $\chi^{2}$ Monte Carlo simulations are shown in Figure \ref{mcchi}, where we have plotted histograms of $\chi^{2}$/DOF for our various sample sizes.  Ideally a peak should occur at $\chi^{2}/DOF = 1.0$ for the $\chi^{2}$ test, but we see that there is significant scatter in the histograms for $n < 30$ (Figure \ref{mcchi}).  Most notable is the $n = 5$ histogram in Figure \ref{mcchi}, which has two peaks at $\chi^{2}/DOF > 10$, much higher than the $\alpha = 0.10$ critical value (i.e., the $\chi^{2}$ value above which 10$\%$ of values fall) of 2.41 (DOF = 1).  Thus for small \emph{n}, the $\chi^{2}$ test tends to overestimate the number of failed/non-Gaussian samples.  It is only for $n \geq 30$ that we see the expected peak of 1.0.  Unfortunately, the majority of the CNOC2 groups have $n_{\rm{members}} < 30$, so use of the $\chi^{2}$ test to classify galaxy group dynamics could result in the false identification of many groups with non-Gaussian velocity distributions.\\
\indent{}The results of the Kolmogorov Monte Carlo simulations are shown in Figure \ref{mcKS}, where we show histograms of the computed $D^{*}$ values and indicate the $\alpha = 0.10$ critical value of 1.224 (DA86) with a dotted vertical line.  Unlike the histograms for the $\chi^{2}$ test, we see no scatter in the $D^{*}$ values, even at $n = 5$.  The histograms of the $D^{*}$ values remain remarkably stable over the sample size range, indicating that the Kolmogorov test is reliable for small \emph{n}.  To determine if our simulations produce the expected $D^{*}$ critical values, we compute these values from our histograms (Figure \ref{mcKS}) and compare them to published values cited in DA86.  For each sample size, we are able to reproduce all of the given critical values for the case where the input distribution parameters, $\mu$ and $\sigma$, are known.\\
\indent{}In Section 2.1, we discussed the use of the $A^{2*}$ values for the A--D test, but this is a modification for the case where the distribution parameters are unknown.  When the input parameters are known \emph{a priori}, as is the case with our Monte Carlo simulations, DA86 state that no modification for the A--D test is needed and one should use the $A^{2}$ values when comparing to critical value tables. The results of the A--D test simulations are shown in Figure \ref{mcAD}, where we plot histograms of the computed $A^{2}$ values and indicate the $\alpha = 0.10$ critical value of 1.933 (DA86) with a dotted vertical line.  The histograms are similar to those of the Kolmogorov test, showing no scatter in the $A^{2}$ values over the entire sample size range.  The stability of the A--D test, even at $n = 5$, supports the claim of DA86 that the statistic is reliable for all $n \geq 5$.  We compute the $\alpha = 0.10$ critical values for the A--D test from the histograms in Figure \ref{mcAD} and find that our values are in agreement with those found in DA86.\\

\newpage
\subsection{Power Comparisons of the Kolmogorov and A--D Tests}
\indent{}Our Monte Carlo simulations indicate that the $\chi^{2}$ test is indeed an unreliable statistic for testing Gaussianity in small sample sizes, as suggested by DA86.  Thus, we can eliminate the use of this test for classification of the galaxy group dynamics.  However, the results of our simulation also show that both the Kolmogorov and A--D statistics are reliable down to $n = 5$.  In order to determine which test is most suitable for distinguishing between Gaussian and non-Gaussian systems, we perform power studies of the Kolmogorov and A--D tests.  Monte Carlo simulations of both tests are performed, using a skewed Gaussian distribution, $F(x_{i})$ \citep{Azzalini}, given by: 
\begin{center}
\begin{equation}
F(x_{i}) = 2\phi(x_{i})\Phi(\alpha_{\mbox{s}}x_{i})
\label{skewG}
\end{equation}
\begin{equation}
\phi(x) = \frac{1}{\sqrt{2\pi}}e^{-\frac{x_{i}^{2}}{2}}
\end{equation}
\end{center}
where $\phi(x)$ is the probability density function of a Gaussian distribution, with $\mu = 0.0$ and $\sigma = 1.0$, $\Phi(x_{i})$ is the CDF of a Gaussian distribution (Equation (\ref{CDF})) and $\alpha_{\mbox{s}}$ is known as the shape or slant parameter.  The value of $\alpha_{\mbox{s}}$ is proportional to the skewness of the Gaussian distribution, with higher values of $\alpha_{\mbox{s}}$ producing more skewed distributions and an $\alpha_{\mbox{s}} = 0.0$ producing a non-skewed Gaussian.\\
\indent{}We draw various sample sizes, $5 \leq n \leq 100$, from a Gaussian random distribution, using 30,000 iterations, and apply varying levels of skewness, $0 \leq \alpha_{\mbox{s}} \leq 5$, to determine the rejection rate of both the Kolmogorov and A--D tests.  The results of our power studies are given in Table \ref{power}, where Column 1 indicates the test used, Column 2 indicates the $\alpha_{\mbox{s}}$ value, and Columns 3 - 9 indicate the percentage of rejection given a specific sample size.  The rejection rates are determined using the 10$\%$ critical values given in DA86.  In order for a test to be considered powerful, the simulations with high $\alpha_{\mbox{s}}$ should have higher rejection rates, since the underlying distributions are increasingly less Gaussian.  From Table \ref{power}, it is clear that for all levels of skewness and all sample sizes, the A--D statistic rejects more objects than the Kolmogorov test.  For the $\alpha_{\mbox{s}} = 0.25$ and $ n \leq 30$ cases the two tests are comparable, but as one increases $\alpha_{\mbox{s}}$, the percentage of rejections for the Kolmogorov test is significantly lower than those of the A--D test.  Looking at the rejection rates for the $\alpha_{\mbox{s}} = 1.0$, which is a heavily skewed Gaussian, it is clear that the Kolmogorov test underestimates the number of non-Gaussian systems.  This is especially evident when one focuses on the $n = 30$ case for the $\alpha_{\mbox{s}} = 1.0$ distribution, which has a 100$\%$ rejection rate for the A--D test, but only an 85$\%$ failure rate for the Kolmogorov.\\
\indent{}The results of our power studies indicate that the Kolmogorov test is much less powerful at detecting departures from Gaussianity than the A--D statistic, which is in agreement with the findings of \citet{Stephens}.  The strongest evidence for this claim is shown in the $\alpha_{\mbox{s}} = 5.0$, a completely non-Gaussian distribution, and $n = 5$ simulation, where the A--D test rejects 100$\%$ of the sample while the Kolmogorov test only rejects 75$\%$ of the sample.

\begin{table}[hbt!]
\begin{center}
\caption{Results of the Power comparisons of the Kolmogorov and A--D tests using various skewed Gaussian distributions\tablenotemark{a}. \label{power}}
\vspace{0.5cm}
\begin{tabular}{ccccccccccccc}
\tableline\tableline
Test & $\alpha_{\mbox{s}}$ & & & & & & & \emph{n} & & & &\\
 & & & & 5 & 10 & 15 & 20 & 25 & 30 & 40 & 50 & 100\\
% $\alpha_{\mbox{s}}$ & \emph{n} & $A^{2}$ \tablenotemark{a} & $D^{*}$ \tablenotemark{a}\\
\tableline
\\
$A^{2}$ & 0.25 & & & 15 & 17 & 19 & 21 & 24 & 27 & 32 & 37 & 64\\
$D^{*}$ & 0.25 & & & 13 & 15 & 16 & 18 & 20 & 22 & 26 & 29 & 46\\
\\
$A^{2}$ & 0.50 & & & 27 & 38 & 48 & 57 & 66 & 74 & 86 & 93 & 100\\
$D^{*}$ & 0.50 & & & 20 & 27 & 34 & 41 & 47 & 53 & 63 & 72 & 95\\
\\
$A^{2}$ & 0.75 & & & 43 & 62 & 77 & 87 & 93 & 96 & 99 & 100 & 100\\
$D^{*}$ & 0.75 & & & 28 & 42 & 55 & 66 & 74 & 81 & 90 & 96 & 100\\
\\
$A^{2}$ & 1.0 & & & 59 & 81 & 92 & 97 & 99 & 100 & 100 & 100 & 100\\
$D^{*}$ & 1.0 & & & 35 & 57 & 72 & 83 & 91 & 85 & 99 & 100 & 100\\
\\
$A^{2}$ & 1.5 & & & 80 & 96 & 99 & 100 & 100 & 100 & 100 & 100 & 100\\
$D^{*}$ & 1.5 & & & 50 & 79 & 91 & 97 & 99 & 100 & 100 & 100 & 100\\
\\
$A^{2}$ & 2.0 & & & 90 & 99 & 100 & 100 & 100 & 100 & 100 & 100 & 100\\
$D^{*}$ & 2.0 & & & 67 & 86 & 97 & 99 & 100 & 100 & 100 & 100 & 100\\
\\
$A^{2}$ & 5.0 & & & 100 & 100 & 100 & 100 & 100 & 100 & 100 & 100 & 100\\
$D^{*}$ & 5.0 & & & 75 & 96 & 100 & 100 & 100 & 100 & 100 & 100 & 100\\ 
\tableline
\end{tabular}
\tablenotetext{a}{This table gives the percentage of rejection based on the 10$\%$ critical values given in DA86.}
\end{center}
\end{table}

\section{Applying the Tests to Data}
\subsection{The Sample}
\indent{}We now apply the $\chi^{2}$, Kolmogorov, and A--D tests to a sample of galaxy groups identified in the second Canadian Network for Observational Cosmology Redshift Survey (CNOC2) in the redshift range of $0.1 < z < 0.6$ \citep{2001ApJ...552..427C}.  The CNOC2 survey observed $\sim 4 \times 10^{4}$ galaxies covering four patches, 1.5 deg$^{2}$ in area, in the $UBVR_{C}I_{C}$ bands down to a limiting magnitude of $R_{C} = 23.0$.  Spectra of more than 6000 galaxies were obtained with the MOS spectrograph on the Canada--France--Hawaii Telescope (CFHT), with 48$\%$ completeness at $R_{C} = 21.5$ \citep{2000ApJS..129..475Y}.\\
\indent{}Over 200 galaxy groups were identified using a friends-of-friends percolation algorithm in the CNOC2 survey \citep{2001ApJ...552..427C}.  \citet{2005MNRAS.358...71W} (hereafter W05) obtained deeper spectroscopy taken with the low dispersion survey spectrograph (LDSS2) on the Magellan telescope and then redefined group membership with more relaxed algorithm parameters than those used by \citet{2001ApJ...552..427C}.  The original search parameters were optimized so that the group-finding algorithm would identify dense, virialized groups, while the W05 sample included looser group populations.  We would like to quantify how many of these groups have more complex velocity distributions, potentially identifying merger products or systems in the early stages of virialization.\\

\subsection{Estimation of Distribution Parameters $\mu$ and $\sigma$}
\begin{figure}[hbt!]
\includegraphics[height=8cm, width = 8cm]{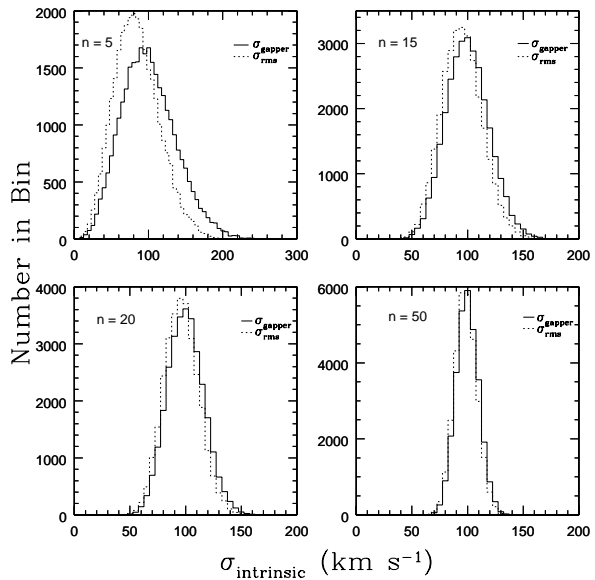}
\caption{Monte Carlo simulations of the Gapper Estimator and the rms dispersion methods for $n = 5$ (top left), $n = 15$ (top right), $n = 20$ (bottom left) and $n = 50$ (bottom right).  For each histogram, we compute the Gapper and rms dispersions using velocities generated from a random Gaussian distribution with input values of $\mu=0.0$ and $\sigma_{\rm{int}} = 100$.  This process was done with 30,000 iterations.  The dotted lines indicate dispersion values computed using the canonical rms standard deviation and the solid lines are values calculated using the Gapper algorithm.  For $n = 5$, the histogram for the rms method is skewed and underestimates the true dispersion by $20\%$. For the same sample size, the peak of the Gapper Estimator histogram is located at the true dispersion value of 100, even though the histogram is also skewed, indicating that the Gapper Estimator is insensitive to outliers.  For the $n = 50$ case, the two methods begin to converge, with both the rms and Gapper algorithms picking out the true dispersion value.}
\label{gapperRMS}
\end{figure}

\indent{}The $\chi^{2}$, Kolmogorov and A--D statistics were developed under the assumption that all of the parameters of the underlying distribution were completely specified.  Modifications to the statistics, with the use of Monte Carlo simulations, have been carried out to allow these tests to be applied to cases where the distribution parameters are not completely known \emph{a priori}, but must be estimated from the data (DA86).\\
\indent{}The parameters required to define a Gaussian distribution are the mean, $\mu$, and the dispersion, $\sigma$.  In the analysis described in Section 3.3, $\mu$ is calculated using the standard mean, and the velocity dispersions are estimated with the Gapper algorithm, given by:
\begin{center}
\begin{equation}
\sigma_{\rm{Gapper}} = \frac{\sqrt{\pi}}{n(n-1)}\sum_{i=1}^{n-1}w_{i}g_{i}
\label{gapper}
\end{equation}
\end{center}
where $w_{i} = i(n-i)$, $g_{i} = x_{i+1} - x_{i}$, here the \emph{ordered} $x_{i}$ values are given by the observed radial velocities of the group members.  For small number statistics, \citet{1990AJ....100...32B} recommend the Gapper Estimator over the canonical rms standard deviation, as this algorithm is insensitive to outliers and thus more accurately reproduces the true dispersion of the system.\\
\indent{}To ensure that the Gapper Estimator is a more reliable method of computing the dispersion of a system, we perform Monte Carlo simulations of both the Gapper algorithm and the canonical rms standard deviation.  We draw various sample sizes ($n = 5, 15, 20, 50$) from a Gaussian random number generator \citep{gsl} with the inputs $\mu = 0.0$ and $\sigma_{\rm{intrinsic}} = 100$ and then compute the dispersion using the two aforementioned methods.\\
\indent{}The results are shown in Figure \ref{gapperRMS}, where we have plotted histograms of the output dispersions for each sample size.  Figure \ref{gapperRMS} shows that for the $n = 5$ case the canonical rms standard deviation underestimates the true dispersion by 25$\%$ and the distribution is heavily skewed to lower values.  Although the distribution of the Gapper Estimator is also skewed, the peak of the distribution occurs at the true dispersion value of 100, indicating that this method is indeed insensitive to outliers.  As we increase the sample size, $n = 15$ and 20, we can see that the rms dispersion continues to underestimate the true velocity dispersion, but also that the two methods begin to converge.  The histogram for the $n=50$ case shows that the rms dispersion and the Gapper Estimator both correctly identify the true dispersion.  Given the small \emph{n} of the CNOC2 groups we choose the Gapper algorithm to estimate the velocity dispersions of the groups.

\subsection{Comparison of Tests}
\subsubsection{Statistical Analysis}
\indent{}The $\chi^{2}$, Kolmogorov and A--D tests are applied to a subset of 62 CNOC2 groups containing at least 10 members per group.  We chose this membership cut based on the result of our $n=5$ $\chi^{2}$ Monte Carlo simulations, as shown in Figure \ref{mcchi}, which indicated that the test was completely unreliable at small \emph{n}.  This cut also helps to minimize the effect of binning.  The potential complications introduced by the choice of bin width are also reduced by using the optimal bin width formula (Equation (\ref{healdwidth})).\\
\indent{}Despite our attempts to minimize the various uncertainties due to binning, we still encounter groups that do not meet the minimum number of bins requirement, $N = 4$, for our application of the $\chi^{2}$ test, with five out of the 62 groups in the sub-sample having $N < 4$.   Thus, we apply the $\chi^{2}$ test to 57 groups ($n \geq 10$ and $N \geq 4$).\\
\indent{}The modification to the \emph{D}-statistic (Equation (\ref{Dstar})) given in Section 2.1 is used for the case when the input distribution parameters are known \emph{a priori} (i.e., Monte Carlo simulations in Section 2.2).  When applying the Kolmogorov test to real data sets, one must estimate the distribution parameters, $\mu$ and $\sigma$, and thus the modification for the \emph{D} value becomes (DA86):
\begin{center}
\begin{equation}
D^{*} = D\left(\sqrt(n) - 0.01 + \frac{0.85}{\sqrt(n)}\right),
\label{Dstar3}
\end{equation}
\end{center}
where \emph{D} is given by Equation (\ref{Dvalue}).  \\
\indent{}Similarly, in Section 2.2 we used the $A^{2}$ values, but for real data sets one must use the $A^{2*}$ values (Equation (\ref{A2star})) in order to properly apply the A--D statistic.  Since, the A--D and Kolmogorov tests do not require binned data, we are able to apply both tests to all 62 groups ($n \geq 10$).\\
\indent{}The results of our analysis are presented in Table \ref{GOFtests}, where Column 2 indicates the number of groups used to perform the specified test and Column 3 indicates the number of groups that failed at the 0.10 significance level.  Our initial findings show that these three tests differed in the number of rejected (non-Gaussian) groups, with a 21$\%$ rejection rate for the $\chi^{2}$ test, a $\sim11\%$ rejection rate for the Kolmogorov test and a $\sim16\%$ rejection rate for the A--D test.\\
\indent{}The \emph{D}-statistic of the Kolmogorov test is often used to test for goodness-of-fit, but both \citet{Stephens} and DA86 do not recommend its use for testing Gaussian distributions, based on its lack of power.  The results of our own power studies, in $\S$2.3, also indicate that the Kolmogorov test lacks power and is unable to detect real departures from Gaussianity.  The relatively low non-Gaussian detection rate in the CNOC2 group sub-sample, with only seven out of the 62 CNOC2 groups failing at the 0.10 significance level, further supports the notion that the Kolmogorov test suffers from under-rejection, suggesting that the Kolmogorov test may not be reliable method for dynamical classification.\\
\begin{table}[hbt!]
\begin{center}
\caption{Results of the $\chi^{2}$, Kolmogorov and A--D tests applied to a sample of CNOC2 groups with $n \geq 10$.\label{GOFtests}}
\vspace{0.5cm}
\begin{tabular}{lcccc}
\tableline\tableline
Test &  Number  & Number of  & Percent of  & Significance\\
         & of Groups & Failed Groups & Failed Groups & Level\\
\tableline
Pearson's $\chi^{2}$ & 57 & 12 & $\sim 21$  & 0.10\\
Kolmogorov & 62 & 7 & $\sim 11$ & 0.10\\
A--D & 62 & 10 & $\sim 16$ & 0.10\tablenotemark{a}\\
\tableline
\end{tabular}
\tablenotetext{a}{The critical value for the A--D test is for case 3, where both $\mu$ and $\sigma$ are unknown.}
\end{center}
\end{table}

\subsubsection{Velocity Distributions}
\begin{center}
\begin{figure}
\includegraphics[width=8cm,height=8cm]{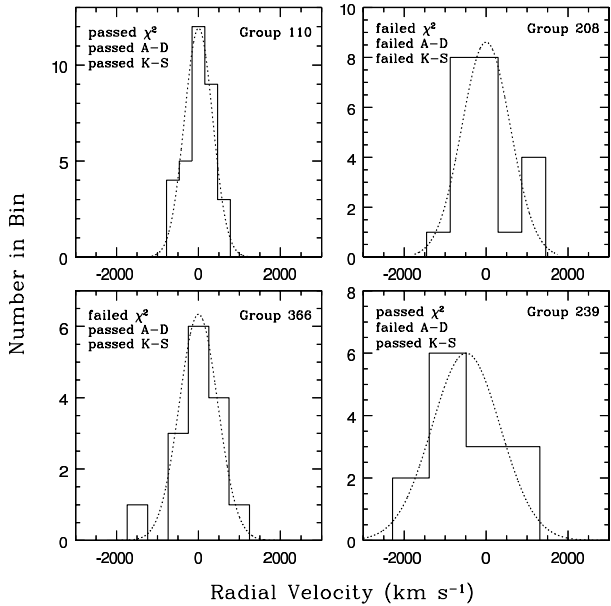}
\caption{Velocity distribution histograms for various CNOC2 groups.  The dotted lines show Gaussian distributions computed using the estimated $\mu$ and $\sigma$ of the group.  The histograms are generated using Heald's optimal bin width (Equation (\ref{healdwidth})), thus the bin widths differ between groups.  Group 110 represents groups that have passed the $\chi^{2}$, Kolmogorov, and A--D tests (at the 10$\%$ level) and are classified as Gaussian.  Group 208 represents groups have failed all three tests and are thus classified as non-Gaussian.  The histogram for Group 208 appears non-Gaussian, showing a double peak.  Group 366 represents groups that have failed the $\chi^{2}$ test but passed the Kolmogorov, and A--D tests.  The histogram for Group 366 shows no obvious departures from the Gaussian distribution, but suffers from binning issues, resulting in rejection by the $\chi^{2}$ test.  Group 239 represents groups that have passed the $\chi^{2}$ and Kolmogorov test, but failed the A--D test.  The histogram for this group appears skewed and has a mean that is slightly shifted from that of the Gaussian distribution.  Despite these features, only the A--D test is able to detect these departures and rejects Group 239 as having an underlying Gaussian distribution.  Note that the Kolmogorov and A--D tests do not require binned data and are computed using ordered data.  Also, note that the \emph{y}-axes differ between plots. }
\label{grphist}
\end{figure}
\end{center}

\indent{}The reliability of the A--D test over the $\chi^{2}$ and Kolmogorov tests is further demonstrated when one looks at the velocity distributions of specific groups.  In Figure \ref{grphist}, we show the velocity distributions of four CNOC2 groups that have either failed the $\chi^{2}$, Kolmogorov, and A--D tests (non-Gaussian groups), passed all three tests (Gaussian groups), failed the $\chi^{2}$ test but passed the Kolmogorov and A--D tests, or have passed the $\chi^{2}$ and Kolmogorov tests but failed the A--D test.  The histograms are made using Heald's optimal bin width (Equation (\ref{healdwidth})) and are over-plotted with a Gaussian distribution generated using the estimated mean and dispersion of the group.  Although the Kolmogorov and A--D tests do not use binned data, we can look at the velocity distributions of groups classified as Gaussian or non-Gaussian by the various tests to see if there are obvious visual departures from normality.\\
\indent{}Group 110 is classified as Gaussian by the $\chi^{2}$, Kolmogorov, and A--D tests and from Figure \ref{grphist} we can see that the shape and mean of the velocity distribution agrees well with the underlying Gaussian distribution.  Group 208 (Figure \ref{grphist}) is a group that has been classified as non-Gaussian by all three tests, and it is evident that the velocity distribution is non-Gaussian, as the histogram shows a double peak.\\
\indent{}While Groups 110 and 208 are examples of systems with obvious Gaussian or non-Gaussian features in their velocity distributions, this distinction is not so clear for several of the CNOC2 groups.  Group 366 represents groups that have failed the $\chi^{2}$ test, but have passed the Kolmogorov and A--D tests.  The histogram for Group 366 (Figure \ref{grphist}), shows no obvious departures from the Gaussian distribution and despite the use of Heald's optimal bin width, the relatively low group membership ($n = 15$) results in binning issues and causes the group to be rejected by the $\chi^{2}$ test.  The Kolmogorov and A--D tests use \emph{ordered} rather than binned data, and thus do not suffer from the issues introduced when binning small sample sizes.\\
\indent{}Group 239 represents groups that have passed the $\chi^{2}$ and Kolmogorov tests, but not the A--D test.  This histogram for this group (Figure \ref{grphist}) appears skewed and and has a mean that is slightly shifted from that of the Gaussian distribution.  Despite these non-Gaussian features, only the A--D test is able to detect these departures and rejects Group 239 as having an underlying Gaussian distribution.  This group highlights the fact that the Kolmogorov test lacks power, as discussed in Section 2.3, and is unable to detect slight departures from Gaussianity.\\

\newpage
\section{Application of the Anderson-Darling Test to the CNOC2 Groups}
\begin{figure}[hbt!]
\centering
\includegraphics[width=8cm,height=8cm]{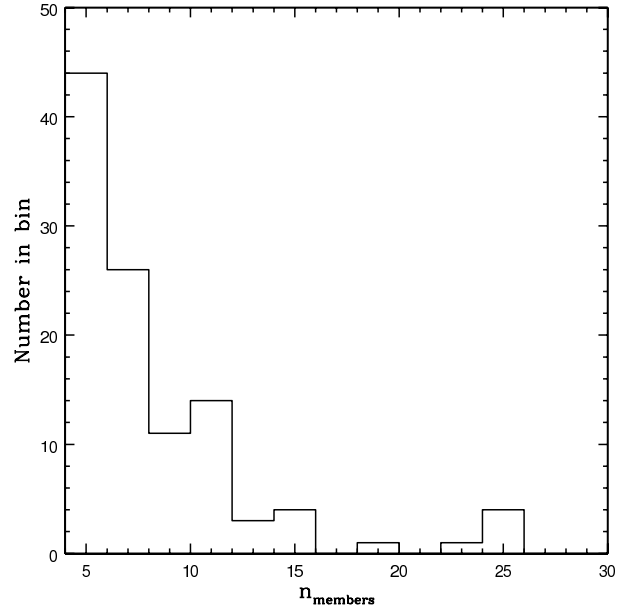}
\caption{Histogram of the number of members per CNOC2 galaxy group in our sample, after a 1 Mpc radius cut and considering only groups with $n\geq 5$.}
\label{nmembers}
\end{figure}

\indent{}Having identified the A--D test as the best statistical tool for galaxy group dynamics analysis, we can now proceed to classify the CNOC2 groups as being either relaxed (Gaussian) or dynamically complex (non-Gaussian) systems.  In our analysis in Section 3, we apply only a minimum membership cut, but proper classification requires a more detailed treatment of the group catalog.  Group members were reallocated by W05, including some galaxies at large group-centric distances to study radial trends.  We apply a 1 Mpc radius cut to each of these re-identified CNOC2 groups.  Figure \ref{nmembers} shows a histogram of the number of group members, after the radius cut, with the majority of groups having 5 - 11 members.\\ 

\indent{}For the distribution parameters, we estimate $\mu$ as the mean group member velocity and $\sigma$ by computing the intrinsic velocity dispersion (W05), obtained by first computing the observed dispersion, given by:
\begin{center}
\begin{equation}
\sigma_{obs} = 1.135\ast\sigma_{Gapper},
\label{sigObs}
\end{equation}
\end{center}
where the factor of 1.135 is a correction applied by W05 for their redshift space cut of 2$\frac{\sigma_{obs}}{c}$.  The next step is to compute the rest-frame velocity dispersion:
\begin{center}
\begin{equation}
\sigma_{rest} = \frac{\sigma_{obs}}{1+ z},
\label{sigrest}
\end{equation}
\end{center}
where \emph{z} is the redshift of the group centroid.  The final step involves removing the measurement uncertainty of each galaxy, $\left<\delta(v)\right>$, from the rest-frame dispersion, as:
\begin{center}
\begin{equation}
\sigma_{int}^{2} =  \sigma_{rest}^{2} - \left<\delta(v)\right>^{2},
\label{sigint}
\end{equation}
\end{center}
where $\left<\delta(v)\right>$ = 142 km s$^{-1}$ for the (LDSS) data and $\left<\delta(v)\right>$ = 103 km s$^{-1}$ for the original CNOC2 data (W05). \\
\indent{}We compute the intrinsic velocity dispersion for all W05 re-identified CNOC2 groups with $n_{\rm{members}} \geq 5$ within 1 Mpc, then classify the groups based on the computed $A^{2*}$ values.  A crucial step in any statistical analysis is to identify the appropriate significance level or critical value used for classification.  The critical points of the A--D test, as with all EDF statistics, change depending on the accuracy with which one knows the input distribution parameters (DA86).  For a Gaussian distribution, there are four cases to be considered:
\begin{itemize}
\item case 0: both $\mu$ and $\sigma^{2}$ are known a priori, i.e., a fully specified distribution;
\item case 1: $\sigma^{2}$ is known and $\mu$ must be estimated;
\item case 2: $\mu$ is known and $\sigma^{2}$ must be estimated, and;
\item case 3: both $\mu$ and $\sigma^{2}$ must be estimated .
\end{itemize}

Each of these cases result in different critical values, which can greatly alter the number of rejections.  For example, the $5\%$ critical value is 2.492 for case 0, 1.087 for case 1, 2.308 for case 2 and 0.752 for case 3 (DA86).  When testing for Gaussianity in data sets, \citet{Stephens} suggests that case 3 is the most practical choice, as distribution parameters are in general estimated and not known \emph{a priori}.  In this situation, there are two approaches one can take in distinguishing between Gaussian and non-Gaussian systems; the first involves comparing the computed $A^{2*}$ values with critical value tables and the other uses the $A^{2*}$ values to directly compute the significance level $\alpha$, which gives the probability that the data comes from an underlying Gaussian distribution.  We choose to follow the latter method and compute $\alpha$ using the formula:
\begin{center}
\begin{equation}
\alpha = a\exp(-A^{2*}/b),
\label{alpha}
\end{equation}
\end{center}
where $a = 3.6789468$ and $b = 0.1749916$, and both factors are determined via Monte Carlo methods \citep{Nelson}.\\
\indent{}Using Equation (\ref{alpha}), we are able to determine the probability of whether or not each CNOC2 group had a Gaussian velocity distribution, classifying all groups with $\alpha < 5\%$ as being non-Gaussian.  The results of the A--D test are given in Table \ref{classAD}, where we see that $\sim$32$\%$ of the CNOC2 groups are classified as non-Gaussian at the $5\%$ significance level and also that the $\overline{n}$, $\overline{z}$, and $\overline{\sigma}$ are similar for the Gaussian and non-Gaussian groups.  Using this classification scheme, we can now investigate specific group properties to determine if there are any obvious trends or differences between the Gaussian and non-Gaussian groups.  The properties of the CNOC2 groups that are classified as dynamically complex (non-Gaussian) are given in Table \ref{nongauss}.  \\
\begin{table}[hbt!]
\begin{center}
\caption{Anderson-Darling Classification of the CNOC2 groups with $n \geq 5$ after the 1 Mpc cut. \label{classAD}}
\vspace{0.5cm}
\begin{tabular}{lcclll}
\tableline\tableline
Classification & $\#$ of groups & $\%$ of all groups & $\bar{n}$ & $\bar{\rm{z}}$ & $\overline{\sigma_{\rm{int}}}$\\
 & & & & & (km s$^{-1}$)\\
\tableline
Gaussian & 72 & $\sim 68\%$ & 9 & 0.30 & 347\\
non-Gaussian & 34 & $\sim 32\%$ & 9 & 0.37 & 327\\
\tableline
\end{tabular}
\end{center}
\end{table}

\begin{table}[hbt!]
\begin{center}
\caption{Properties of non-Gaussian CNOC2 Groups. \label{nongauss}}
\vspace{0.5cm}
\begin{tabular}{lllllll}
\tableline\tableline
group ID &  $n$\tablenotemark{a} & $\sigma_{obs}$ & $\sigma_{rest}$ & $\sigma_{int}$ & $A^{2*}$ & $\alpha$\\
 & & (km s$^{-1}$) & (km s$^{-1}$) & (km s$^{-1}$) & & \\
\tableline
24 & 10 & 151 & 111 & 42 & 13.4 & 2.69e-33\\
29 & 9 & 465 & 338 & 322 & 0.780 & 0.0426\\
30 & 11 & 428 & 307 & 289 & 1.18 & 0.00424\\
32 & 8 & 755 & 542 & 532 & 0.945 & 0.0166\\
33 & 6 & 235 & 167 & 126 & 1.61 & 0.000375\\
34 & 6 & 254 & 173 & 134 & 1.67 & 0.000271\\
38 & 16 & 1208 & 800 & 793 & 1.66 & 0.00940 \\
120 & 6 & 248 & 200 & 171 & 0.752 & 0.0499\\
127 & 6 & 180 & 137 & 91 & 1.18 & 0.00427\\
128 & 5 & 586 & 445 & 433 & 1.22 & 0.00342\\
129 & 5 & 371 & 282 & 262 & 0.813 & 0.0353\\
132 & 8 & 542 & 399 & 382 & 1.31 & 0.00202\\
135 & 7 & 379 & 271 & 251 & 1.03 & 0.0105\\
138 & 23 & 1064 & 740 & 731 & 0.973 & 0.0142\\
139 & 10 & 363 & 252 & 226 & 1.39 & 0.00128\\
140 & 5 & 219 & 149 & 100 & 1.18 & 0.00432\\
202 & 5 & 423 & 356 & 341 & 0.836 & 0.0309\\
211 & 8 & 234 & 184 & 153 & 0.892 & 0.0225\\
213 & 7 & 581 & 446 & 434 & 0.763 & 0.0469\\
218 & 5 & 443 & 338 & 322 & 0.880 & 0.241\\
221 & 6 & 171 & 126 & 73.1 & 2.37 & 4.863-06\\
226 & 25 & 1159 & 853 & 847 & 1.04 & 0.00940\\
230 & 10 & 212 & 153 & 113 & 1.25 & 0.00298\\
233 & 7 & 795 & 568 & 559 & 1.57 & 0.000463\\
241 & 6 & 316 & 222 & 197 & 0.757 & 0.0487\\
244 & 15 & 356 & 242 & 211 & 1.35 & 0.00168\\
312 & 8 & 424 & 344 & 328 & 1.14 & 0.00545\\
336 & 5 & 965 & 708 & 700 & 0.947 & 0.0164\\
338 & 9 & 503 & 367 & 352 & 0.822 & 0.0336\\
346 & 26 & 613 & 446 & 434 & 1.26 & 0.00277\\
351 & 6 & 243 & 176 & 143 & 0.814 & 0.0352\\
357 & 7 & 234 & 168 & 133 & 1.80 & 0.000128\\
358 & 11 & 409 & 294 & 275 & 0.776 & 0.0436\\
360 & 7 & 878 & 631& 622 & 0.831 & 0.0319\\
\tableline
\end{tabular}
\tablenotetext{a}{Members after 1 Mpc radius cut.}
\end{center}
\end{table}

\begin{figure}[hbt!]
\centering
\includegraphics[width = 8cm, height=8cm]{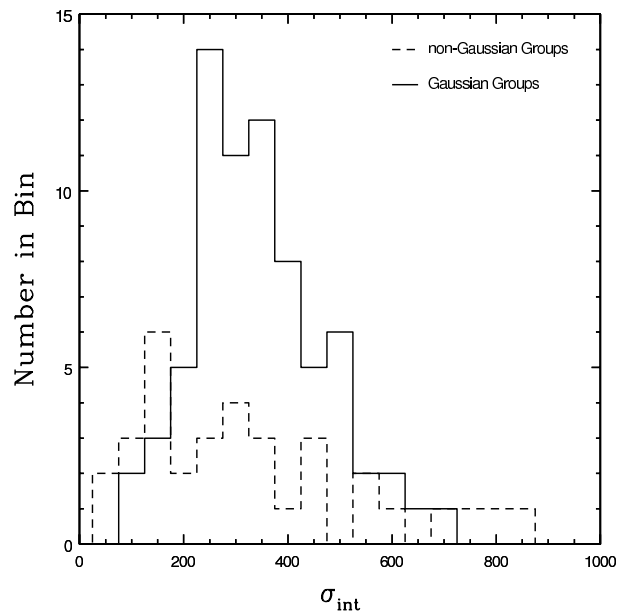}
\caption{Histogram of the computed intrinsic velocity dispersions ($\sigma_{int}$) for CNOC2 galaxy group, with $n \geq 5$ after a 1 Mpc radius cut, which have been classified as either Gaussian (solid line) or non-Gaussian(dashed line) by the A-D test.}
\label{g_nong_sigma}
\end{figure}

\begin{figure}[hbt!]
\centering
\includegraphics[width = 8cm, height=8cm]{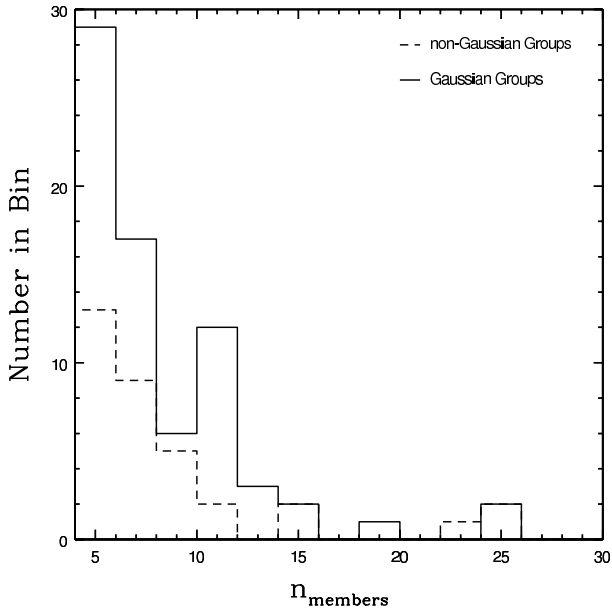}
\caption{Histogram of the number of members per CNOC2 galaxy group, with $n \geq 5$ after a 1 Mpc radius cut, which have been classified as either Gaussian (solid line) or non-Gaussian (dashed line) by the A-D test.}
\label{g_nong_n}
\end{figure}

\indent{}The intrinsic velocity dispersions of the non-Gaussian groups, shown in Figure \ref{g_nong_sigma} show no obvious trend, with the values ranging from $\sim$40 to 850 km s$^{-1}$, so the more dynamically complex systems are not restricted to low or high velocity dispersions. Similarly, the intrinsic velocity dispersions of the Gaussian groups (Figure \ref{g_nong_sigma}) also cover a wide range in values, $\sim$75 - 725 km s$^{-1}$.  Figure \ref{g_nong_n} shows a histogram of the number of members in the CNOC2 groups, with the solid line representing the classified Gaussian groups and the dashed line indicating the non-Gaussian groups.  Here we also see a wide range in the number of members for the Gaussian and non-Gaussian groups, suggesting that the A--D test is not biased towards a specific sample size.  

\section{Velocity Dispersion Profiles}
\begin{figure}[hbt!]
\centering
\includegraphics[width=8.5cm,height=9.5cm]{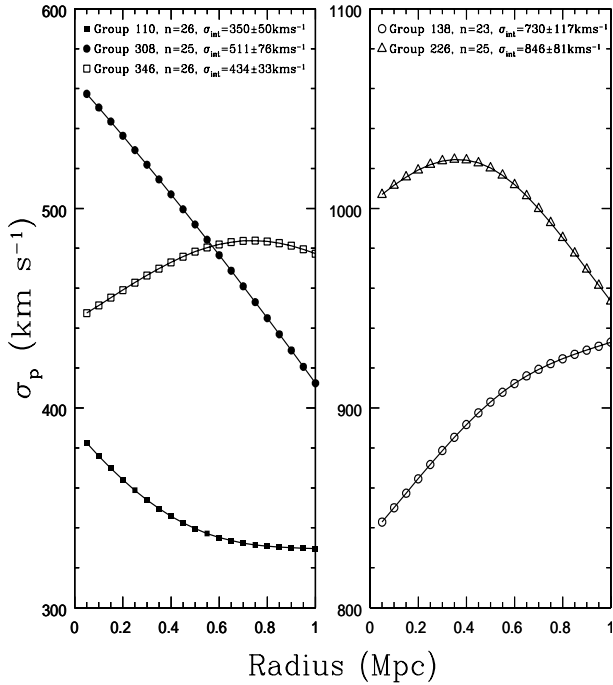}
\caption{VDPs for the CNOC2 groups with $n > 20$ after a 1Mpc radius cut and using a constant window width of 0.35 Mpc.  The open symbols indicate groups classified as non-Gaussian and the closed symbols indicate those classified as Gaussian.}
\label{VDPs}
\end{figure}

\indent{}One way to investigate the differences between relaxed and dynamically complex galaxy groups is to study their VDPs.  If a group classified as non-Gaussian is in fact dynamically more complex than a Gaussian one, then the corresponding VDP may exhibit different features from those of relaxed systems.  In their analysis of merging clusters, \citet{1996ApJ...472...46M} find that radially increasing VDPs indicate significant galaxy merging in the cluster core. \citet{1996ApJ...457...61G} also suggest that the presence of neighboring clusters resulted in a VDP with an initially flat profile, which then increased significantly at larger radii.\\
\indent{}We follow the method outlined in \citet{2006A&A...448..155B} to generate VDPs for the CNOC2 groups in which the radial velocities are binned with an exponentially weighted moving window.  The window function is given by:

\begin{center}
 \begin{equation}
 w_{i}(R) = \frac{1}{\sigma_{R}}\exp\left[\frac{(R-R_{i})^{2}}{2\sigma_{R}^{2}}\right],
\label{vdpwindow}
\end{equation}
\end{center}
where $\sigma_{R}$ is the width of the window, which can be constant or a function of radius \emph{R}, and the $R_{i}$'s are the radial positions of the members of the system.  The projected velocity dispersions are then defined as:
\begin{center}
 \begin{equation}
  \sigma_{p}(R) = \sqrt{\frac{\sum_{i}w_{i}(R)(x_{i} - \bar{x})^{2}}{{\sum_{i}w_{i}(R)}}}
\label{sigmap}
 \end{equation}
\end{center}
where the $x_{i}$'s are the radial velocities and $\bar{x}$ is the mean velocity of the system.\\
\indent{}This ``moving window'' prescription for computing VDPs takes into account the contribution of every radial velocity measurement at each value of \emph{R}.  It also removes the restriction of computing binned projected velocity dispersion.  Instead, a smoothed profile can be generated, since the projected dispersions can be computed at any radius, not just at the radii corresponding to the observed velocities.\\
\indent{}In order to use this method to probe the dynamics of a system, one must be careful with the choice of window width, $\sigma_{R}$, as a window that is too large can wash out real features and, a window that is too small tends to add spurious features in the profile.  VDPs of systems with very small \emph{n} may also contain unphysical features, similar to profiles made with small window widths.  Since the projected velocity dispersions are computed using weighted values of every velocity measurement in the data, any large individual deviations from the mean can alter the overall shape of the profile.  This effect is significantly more pronounced in small samples, as there are not enough data points to counteract or outweigh the effects of an outlier.  With testing, a window width of 0.35 Mpc, approximately one-third the value of the maximum radius is selected.  We also enforce a minimum group membership of 20 members, after our 1 Mpc radius cut, to ensure that any visible trend in the VDPs is not a result of outliers.\\
\indent{}There are five CNOC2 groups that meet our minimum group membership criteria ($n \geq 20$), two of which are classified by the A--D test as Gaussian, Groups 110 ($n = 26$) and 308 ($n = 25$) and three as non-Gaussian systems, Groups 138 ($n =  23$), 226 ($n = 25$), and 346 ($n = 26$).  The VDPs for these groups are shown in Figure \ref{VDPs}, where the Gaussian groups are shown with filled symbols and the non-Gaussian groups with open symbols.  Both Gaussian groups have decreasing profiles, while two of the three non-Gaussian groups (138 and 346) have increasing profiles.  The VDP for Group 226 (Figure \ref{VDPs}) does not exhibit the same overall trend as the other non-Gaussian groups, as the profile increases initially but turns over at roughly 0.4 Mpc.  Although the profile for Group 226 does not continually increase, it does show distinct features from the profiles of the Gaussian groups.  It is impossible to make general statements on the overall shape of all Gaussian or non-Gaussian groups based on these five groups alone, but the results do support our claim that galaxy groups classified as non-Gaussian by the A--D test are dynamically complex systems.  We are unable to determine by the VDPs alone whether these non-Gaussian groups are indeed undergoing a merger, but the differences between Gaussian and non-Gaussian group profiles do suggest that these two types of systems are dynamically distinct.\\
\indent{}A closer inspection of the individual profiles of the Gaussian groups suggests another interesting result.  The profile for Group 110 shows an initial decrease with an eventual flattening of the profile towards the outer radius, a general trend that was observed in clusters by \citet{1996ApJ...457...61G}.  The VDP of Group 308 does not exhibit the same trend of a flattened profile, continually decreasing towards the edge of the group.  

\section{Summary}
\indent{}Analysis of galaxy group dynamics requires the use of tools that are reliable and powerful for small sample sizes.  We have applied three goodness-of-fit tests, the $\chi^{2}$, Kolmogorov, and A--D tests, to a subset of the CNOC2 groups in order to determine which test can best classify galaxy group dynamics.  Based on our initial application of the aforementioned tests and on the results of our Monte Carlo simulations and power studies, we conclude that the A--D test is the most reliable statistic to distinguish between relaxed (Gaussian) and dynamically complex (non-Gaussian) groups.\\
\indent{}The results of our Monte Carlo Simulations for the Gapper Estimator (Equation (\ref{gapper})) and rms dispersion calculations indicate that for small sample size, $n < 50$, the Gapper algorithm is a more accurate estimate of the true velocity dispersion, which is in agreement with \citet{1990AJ....100...32B}.\\
\indent{}We then apply the A--D test to all CNOC2 groups with $n_{\rm{members}} \geq 5$, after a 1.0 Mpc radius cut, using the mean velocity of the group members as the estimated $\mu$ and the intrinsic velocity dispersion (Equation (\ref{sigint})) as the estimated $\sigma$.  The groups are then classified as being in either a relaxed (Gaussian) or complex (non-Gaussian) dynamical system, based on their computed significance levels.  The results of our analysis indicate that 34 of the 106, or  $\sim32\%$, of the sample of CNOC2 groups were classified as non-Gaussian.\\
\indent{}To investigate our claim that classified non-Gaussian groups are indeed dynamically more complex than Gaussian ones, we look at the VDPs of five CNOC2 groups with $n_{\rm{members}} \geq 20$.  Analysis of the resulting profiles indicates that;

\begin{enumerate}
 \item the profiles of the two Gaussian groups (110 and 308) show a declining velocity dispersion with radius;
 \item two non-Gaussian groups (138 and 346) have rising profiles, a possible signature of mergers \citep{1996ApJ...472...46M}, and;
 \item the profile of Group 110 flattens towards larger radii, a trend observed by \citet{1996ApJ...457...61G} in galaxy clusters.
\end{enumerate}
The VDPs of the Gaussian and non-Gaussian groups are distinct, supporting our claim that the classified non-Gaussian groups are dynamically different from the Gaussian systems.  \\
\indent{}We intend to investigate other observed group properties, such as morphology, colour and star formation rates, for correlations between the dynamical state of a group and its properties.  We also plan to study the dynamics of X-ray selected groups (J. Connelly, et al, in preparation; \citet{Finoguenov})  in the CNOC2 fields.  X-ray bright groups are of particular interest, as only groups in a relaxed dynamical state are expected produce extended X-ray emissions.  The application of the A--D test to these groups will allow us to test this hypothesis.  Future work will also include a detailed study of simulated galaxy groups in order to understand the relationship between velocity dispersion and mass for systems in different dynamical states.  Simulations will also allow us to quantify how projection and contamination by interloping galaxies affect our measured velocity distributions.\\
\indent{}We conclude that the A--D goodness-of-fit test is a reliable statistical tool.  Although we have used this test to determine departures from normality, its application is not restricted to Gaussian distributions and can be used with many continuous or discrete distributions.  Not only is this test reliable and powerful, especially when dealing with small sample sizes, but its application is simple and has the potential to be useful is many other areas of astronomy.

\acknowledgments
We thank the CNOC2 team for the use of their unpublished redshifts. L.C.P. acknowledges support from a NSERC Discovery Grant.

\bibliographystyle{apj}
\bibliography{annie_stats_paper}

\end{document}